\documentclass{article}
\usepackage{arxiv}
\usepackage{tabularx}
\usepackage{amsmath}
\usepackage[utf8]{inputenc} % allow utf-8 input
\usepackage[T1]{fontenc}    % use 8-bit T1 fonts
\usepackage{hyperref}       % hyperlinks
\usepackage{url}            % simple URL typesetting
\usepackage{booktabs}       % professional-quality tables
\usepackage{amsfonts}       % blackboard math symbols
\usepackage{nicefrac}       % compact symbols for 1/2, etc.
\usepackage{microtype}      % microtypography
\usepackage{lipsum}
\usepackage{graphicx}
\graphicspath{ {./images/} }

\begin{document}
\title{Does Extending Polling Hours Compensate for Bomb Threats?\\
\Large{Evidence from the 2024 Election in Georgia, USA} }

\author{
 Sequoia Andrade \\
  Department of Statistics\\
  University of California\\
  Berkeley, CA 94720-3860 , USA \\
  \texttt{srandrade@berkeley.edu} \\
  %% examples of more authors
   \And
 Philip Stark \\
  Department of Statistics\\
  University of California\\
  Berkeley, CA 94720-3860 , USA \\
  \texttt{pbstark@berkeley.edu} \\
}

\maketitle
\begin{abstract}
At least 227~bomb threats against polling places and tabulation centers were received on the day of the 2024 US presidential election.
    Threats disrupted voting in the swing states of Georgia, Arizona, and Pennsylvania while law enforcement swept polling places.
    No bombs were found.
    Roughly a dozen `credible' threats were received in Georgia in DeKalb and Fulton counties, interrupting voting at eleven polling locations; there were dozens of `non-credible' threats in addition. 
    To remediate the effect of the temporary closures of polling places, courts ordered some polling places to remain open late.
    Nonparametric statistical tests using turnout data from 2020 and 2024 show that this remedy may have been inadequate: in DeKalb, 2024 in-person turnout in precincts closed by threats 
    relative to 2020 turnout in the same precincts was suppressed compared to other DeKalb polling places ($P \approx 0.01$). 
    In Fulton, there is no statistical evidence 
    that in-person voting was suppressed in precincts closed by bomb threats. 
\end{abstract}

% keywords can be removed

\section{Introduction: Legal and Illegal Disenfranchisement in the USA}
% lead with importance of election integretity - info about historical voter intimidation
% summarize what happened in georgia on election day 
% lead up to in this paper

The US Voting Rights Act (VRA) of 1965 banned many discriminatory voting practices throughout the US, such as literacy tests \cite{woodsVoterSuppressionGeorgia2020, ottoboniElectionIntegrityElectronic2019a}. 
These protections against disenfranchisement were essentially eliminated when the 2013 \textit{Shelby County v.\ Holder} ruling found certain provisions of the VRA to be unconstitutional \cite{woodsVoterSuppressionGeorgia2020, ottoboniElectionIntegrityElectronic2019a}. 
New laws passed in the wake of \textit{Shelby County v.\ Holder} may disenfranchise some eligible voters.

For instance, in the state of Georgia (`Georgia' henceforth), voters are required
to use unreliable, untrustworthy ballot-marking devices to vote in person \cite{starkWhenAuditsRecounts2024a,ottoboniElectionIntegrityElectronic2019a,halderman21,CISA22};
polling sites have been closed \cite{wilderVoterSuppression20202025};
a 2021 law tightens identification requirements
\cite{wilderVoterSuppression20202025, woodsVoterSuppressionGeorgia2020}, shortens the window to request an absentee ballot, reduces the number of ballot drop boxes, tightens rules for counting provisional ballots, limits the use of mobile voting stations to `declared disasters,' and prohibits providing food or drink (even water) to voters waiting to vote \cite{sb202};%\footnote{%
%\url{https://www.fultoncountyga.gov/inside-fulton-county/fulton-county-departments/registration-and-elections/sb-202-changes}, last visited 13~July 2025.
%}
voter rolls have been purged \cite{wilderVoterSuppression20202025,swiftUnravelingRiseMass2024}; registration has been made more difficult \cite{woodsVoterSuppressionGeorgia2020};
and there have been mass challenges to voter registration \cite{swiftUnravelingRiseMass2024}. 
In 2025, President Trump issued an executive order that requires specific forms of identification to verify citizenship, introduces restrictions on postal ballot acceptance, strengthens the federal government's authority over elections, and targets election crimes, such as voter intimidation \cite{execOrder2025}. 

These measures may reduce illegal voting---but illegal voting already seems to be quite 
rare in the USA, on the order of $10^{-6}$ or less~\cite{kamarck2024how}.
Hence, the marginal risk that these measures will disenfranchise eligible 
voters may be far larger than any marginal reduction of fraudulent voting they may entail.
The tradeoff involves moral, legal, and political considerations.

Voter intimidation at the polls has increased since 2020.
For instance, some poll watchers intimidate voters by openly carrying firearms, which is legal in many jurisdictions \cite{becker2024, morales2023, vesoulis2022vigilantes}.
 Intimidation of election workers is also increasing \cite{becker2024}: in one survey, one out of six workers in local elections in 2022 reported being threatened \cite{survey2022}.

Illegal acts also disenfranchise voters.
Bomb threats disenfranchise voters through intimidation and denial of service.
In 2024, at least 227~polling places, election offices, and tabulation centers received threats \cite{howard2024}.\footnote{%
    See also \url{https://www.fbi.gov/news/press-releases/fbi-statement-on-bomb-threats-to-polling-locations}, \url{https://www.npr.org/2024/11/06/nx-s1-5181834/election-day-voting-bomb-threats},
    last accessed 13~July 2025.
} 
Threats to polling places were concentrated in Arizona, Pennsylvania, and Georgia---swing states that sometimes 
have a Democratic majority and sometimes a Republican majority.
These states have particular leverage on the outcome of US presidential elections because, like most states, they cast all their electoral college votes (11, 19, and 16, respectively) for the state's popular winner.

The polling places in Georgia that received bomb threats were primarily in DeKalb and Fulton Counties, in the predominantly Democratic Atlanta metropolitan area.
Fourty-five percent of Fulton residents 
and fifty-three percent of DeKalb residents are Black \cite{census}. 
Dozens of threats were not considered `credible,' but
the credible threats led to evacuations to sweep for bombs at 6
polling places in DeKalb and 5 in Fulton.
(`Credible' amounts to `resulted in a poll closure' below.)
Polls were closed from 10~minutes to approximately an hour~\cite{pbs2024voting, dekalb2024deklab}.
% \footnote{%
% \url{https://www.DeKalbcountyga.gov/news/DeKalb-county-responds-multiple-bomb-threats}; \url{https://www.pbs.org/newshour/politics/voting-hours-extended-at-five-polling-places-in-georgias-fulton-county-disrupted-by-bomb-threats},
%  last accessed 13~July 2025.
% } 
Judges ordered some polling places to remain open from 10 to 89~minutes beyond the official closing time of 19:00 to compensate~\cite{docket_dekalb, docket_fulton}.
%\footnote{%
%\url{https://www.democracydocket.com/news-alerts/these-election-day-polling-places-will-stay-open-late/},  last accessed 13~July 2025.
%}

There are claims that the bomb threats had minimal impact \cite{howard2024}, but we are not aware of any empirical basis for that claim.
This paper uses turnout data from 2020 and 2024 and permutation tests to assess whether there was an impact on turnout, despite the extended voting hours in polling places that were closed.
We conclude that in DeKalb County, there was an impact; however,
the situation in Fulton is less clear, in part because there were many additional threats deemed `non-credible' where polling places remained open and because of complex absentee voting behavior during the 2020 election. 
%in part because many threats in Fulton were deemed `not credible' so the polling places remained open.
%There are other interesting questions about the bomb threats, such as why these locations were chosen  and whether the suppression could have changed the electoral outcome; we do not address those questions here.
Section~\ref{sec:background} provides brief context on elections in the US and Georgia specifically.
Section~\ref{sec:2024} discusses the 2024 election and the credible Fulton and DeKalb
bomb threats.
Sections~\ref{sec:methods} and \ref{sec:results} discuss methodology, data, and results.
Section~\ref{sec:discuss} discusses the implications. 

\section{The Context of the 2024 Election}
\label{sec:background}

%We summarize some aspects of elections in the US and Georgia specifically.

\subsection{US Elections} % ?
% Election types: General vs Midterm
% Voting types: early, day off, absentee by mail
% Covid-19 context: Democratic increase in mail in methods
% Stop the steal context?

There are three categories of political elections in the US: general elections, midterm elections, and special elections. 
General elections occur on the first Tuesday of November every four years and allow voters to vote for President, the House of Representatives, one-third of the Senate, and other local positions (e.g., mayor) and propositions (e.g., state bills). 
Midterm elections occur every two years and include some seats in the House of Representatives, one-third of the Senate, and local positions and propositions. 
Special elections occur as needed, such as when a politician vacates an office (steps down, is recalled, or takes a different position in government).
Typically, far more voters participate in general elections than in midterm or special elections. 
The November 2024 election was a general election: voters voted for the President in addition to other ballot items.

Each US state administers its own elections, including elections for federal offices. 
Depending on the jurisdiction, voters generally can choose among voting on election day in person, voting early in person, or voting by mail (absentee, VBM). 
Those who show up to vote in person but are not listed in the pollbook as registered voters generally are allowed to cast a provisional ballot, which may or may not be included in the tally, depending on whether the voter is later determined to be eligible. 
States have different rules
regarding voting methods, in particular for voting absentee \cite{DemocracyMapsAbsentee}.
%Fourteen states require eligible voters to have a valid excuse (e.g., disabled and cannot make it to the poll site) to obtain an absentee ballot and vote by mail, while twenty-eight states do not require an excuse and eight states allow all elections to be vote by mail \cite{DemocracyMapsAbsentee}. 

\subsection{Georgia and Election Integrity} % ?
% Electric voting problems/ Day of voting hours

Georgia voters can vote in person on election day only at the polling location for their precinct,
which is assigned based on their home address. 
On election day, polls are open from 7:00--19:00 and stay open until everyone in line by 19:00 has voted. 

%The state of Georgia, including DeKalb and Fulton counties in particular, is notorious for having very long lines at certain polling locations \cite{wilderVoterSuppression20202025}. 

% stop the steal
% audits
% recounts
% Curling v. Raffensperger
% "When audits and Recounts are a Distraction"
% Gov. Kemp's letter to Raffensperger
% 

% https://www.columbian.com/news/2025/jan/22/votes-for-mr-potato-head-and-fictional-characters-given-to-a-real-candidate-in-south-georgia-race/ 
% https://apnews.com/article/2nd-georgia-county-find-uncounted-votes-018eac6ac24733d63d356ee76f485530
% https://billmoyers.com/story/georgias-hand-count-of-2020-ballots-was-no-risk-limiting-audit/
% https://myemail.constantcontact.com/Mickey-Mouse-and-Friends-Hijack-Georgia-s-Vote-Count-in-Sumter-County.html?soid=1109272168263&aid=MwpBcNL3PgU

Many believe that the 2020 presidential election was stolen \cite{bartels2023house, becker2024}.
Georgia Secretary of State Brad Raffensperger was pressured to overturn the 2020 election results over claims of fraud \cite{miller2023}. 
Even before 2020, Raffensperger was sued over the security and reliability of Georgia's electronic voting systems \cite{curlingVraffensperger}.
The Georgia Republican party has sought to ban Raffensperger from running for office again as
a Republican~\cite{amy2025georgia}.

Deficiencies of election audits in Georgia are well documented, despite claims by
Raffensperger that an audit of the 2020 election proved that ``no votes were flipped.''~\cite{Raffensperger2020historic}
% \footnote{%
% \url{https://sos.ga.gov/news/historic-first-statewide-audit-paper-ballots-upholds-result-presidential-race}, last accessed 13~July 2025.
% }
Serious problems with the election and audit have been found, such as thousands of votes missing in multiple counties~\cite{brumback2020second};
ballots included in the machine tally and machine recount multiple 
times \cite{starkWhenAuditsRecounts2024a}; and audit tally batches omitted from the 
audit totals \cite{starkWhenAuditsRecounts2024a}.
After the 2024 election, Georgia claimed to have proved that only 87~votes were misinterpreted statewide~\cite{wilson2024state},
but it was later shown that
over a thousand write-in votes had been attributed to the wrong candidate~\cite{niesse2025votes}.
Fulton County is known for having relatively few polling places, long lines, and slow mail-in ballot processing times, especially in majority Black communities \cite{wilderVoterSuppression20202025}.
These conditions all contribute to disenfranchisement.

\section{The 2024 General Election and Bomb Threats}
\label{sec:2024}

The 2024 general election was on 5 November 2024.
The presidential race between Kamala Harris (Democratic) and Donald Trump (Republican) was projected to be tight.\footnote{%
\url{https://www.nytimes.com/interactive/2024/us/elections/polls-president.html}, last accessed 13~July 2025.
} 
%\subsection{The Bomb Threats} 
In the Atlanta, Georgia, metropolitan area, bomb threats primarily targeted the largely Democratic DeKalb and Fulton counties, each of which received credible threats in a handful precincts. 
%where and when were the bomb threats?
%how long before the buildings were declared safe?
%Which court(s) ordered the polls to stay open late, and for how long?

There were eight credible bomb threats in DeKalb County, of which six were to active polling sites, which
were closed and searched by police prior to reopening. 
The polling locations, their corresponding precincts, hours closed to sweep the premises, duration of closure, and extended polling hours are in Table~\ref{tab:DeKalb}. 

The first location to close was New Bethel African Methodist Episcopal Church around 17:15, followed shortly by four others.
The Briarwood Recreation Center received a threat hours later at 19:45, when only voters who had queued before 19:00 remained.

The Democratic National Committee and the Democratic Party of Georgia sued the DeKalb County Department of Registration and Elections to extend poll closing times at locations closed due to bomb threats~\cite{docket_dekalb}.
The DeKalb County Law Department had already decided to extend polling hours, so the suit was withdrawn~\cite{dekalb2024police}.
The extensions ranged from thirty-eight minutes to over an hour.

\begin{table}
    \centering
    \caption{DeKalb polling locations closed by `credible' bomb threats, corresponding precincts, threat closure time and duration, and extended polling hours. `N/A' indicates we were unable to find the data.}
    \begin{tabular}{|l|l|c|c|l|}
    \hline
        Location & Precinct & Closed & Duration & Extended \\
        \hline
New Bethel AME Church & Rockbridge Road & 
        17:15--18:06 & 21m & 19:38\\
New Life Community Center & Flat Shoals & 
17:30--18:35 & 65m & 19:38 \\
North DeKalb Senior Center & Chamblee & 
17:30--18:20 & 50m &  20:29\\
Reid H.\ Cofer Library & Tucker Library & 
 17:25--18:20 & 65m & 19:38 \\
William C. Brown Library & Wesley Chapel Library &
17:56--18:08 & 12m & 20:03\\
Briarwood Recreation Center & Briarwood & N/A & N/A & 20:22\\ 
         \hline 
    \end{tabular}
    \label{tab:DeKalb}
\end{table}

Bomb threats to Fulton County polling places started around 8:00. 
Thirty-two Fulton precincts received threats, of which five were deemed credible enough to warrant a response~\cite{pbs2024voting, waddick2024more}.
% \footnote{%
% \url{https://www.pbs.org/newshour/politics/voting-hours-extended-at-five-polling-places-in-georgias-fulton-county-disrupted-by-bomb-threats},
% \url{https://www.usatoday.com/story/news/politics/elections/2024/11/05/georgia-election-bomb-threats/76082394007/},
%  last accessed 13~July 2025.
% }
Locations with credible threats were either evacuated and swept by police before re-opening, or just swept.
Polling locations and corresponding precincts that were temporarily closed by threats are in Table~\ref{tab:fulton};
less information is available for the Fulton closures than for the DeKalb closures.

Two locations in Union City---Etris-Darnell Community Center and Gullatt Elementary School---received threats around 8:15 and were evacuated for about 30 minutes~\cite{jones2024live}.
The Southwest Arts Center was evacuated briefly~\cite{raymond2024what}.
Northwood Elementary School received a threat later in the day: officers responded to a dispatch call of a `hydrogen bomb' at 12:03~\cite{appen2024police}.
Little has been published about the Lake Forest Elementary threat, but the closure seems to have been brief, since polls were kept open only ten extra minutes. 

%Obviously the bomb threats resulted in disruption to the voting process due to evacuations and temporary closures. 
%In an effort to remediate the disruptions from evacuations and temporary closures, polling locations that were temporarily closed extended their hours. 
A law suit was filed by the Democratic National Committee and the Democratic Party of Georgia against the Fulton County Department of Registration and Elections to extend polling hours at the Etris-Darnell Community Center and C.H.\ Gullatt Elementary School~\cite{docket_fulton}.
The court ordered that the polls be kept open beyond 19:00.
Later, the Fulton County Department of Registration and Elections modified the order to include all five affected locations, which  were kept open an extra 10 to 45 minutes.

\begin{table}[h]
    \centering
    \caption{Fulton polling locations closed by bomb threats, corresponding precincts, and closure time extended past 19:00.
    `N/A' means we were unable to find the data.}
    \begin{tabularx}{\textwidth}{|l|X|c|c|l|}
    \hline
        Location & Precinct & Closed & Duration & Extended \\
        \hline
         Etris-Darnell Community Center  & UC02 A/B/E/F/G & N/A & $\approx$30m & 19:45\\
        C.H. Gullatt Elementary School & UC02 C/D & 8:15-8:53 & 38m & 19:15\\
        Southwest Arts Center & SC02 \& SC02A, SC32 & N/A & N/A & 19:43\\
        Northwood Elementary School & RW06, RW13 & 12:11--12:47 &
        36m & 19:45\\
        Lake Forest Elementary School & SS03, SS03A, SS07A/B/C/D & N/A & N/A & 19:10\\
         \hline 
    \end{tabularx}
    \label{tab:fulton}
\end{table}

\section{Methodology}
\label{sec:methods}
This section explains the data on mode of voting
and the statistical approach to testing differences in in-person voter turnout. 
The methods test whether precincts with bomb threats had relatively lower in-person voter turnout than precincts without bomb threats.
Here, `relatively' compares 2024 to 2022, allowing each precinct to serve as its own control group.
\subsection{Data Acquisition}
% - threat data
% - election data for 2024, 2020, 2016
% + 2024: webscrapped securitary of state website to get vote method breakdown https://results.sos.ga.gov/results/public/Georgia/elections/2024NovGen 
% + 2022: clarity elections (fulton, DeKalb (also from their website)
% + 2020, 2016: clarity elections - note only has per race 

% Add note on data disappearing from the website afterwards - perhaps check on wayback machine

For our statistical analysis, we assembled a data set of turnout by mode of voting for each precinct in the 2024 and 2020 general elections.
(As noted below, some precincts existed in 2020 but not 2024, and vice versa.)
Within days of the election, precinct-level voting data were available on the Georgia Secretary of State website.\footnote{%
\url{https://results.sos.ga.gov/results/public/Georgia/elections/2024NovGen}, last accessed 13~July 2025.
}
On the evening of November 6th, the day after the election, we used a programmatic web scraper to extract the number of absentee votes, early votes, election-day in-person votes, and provisional votes for each precinct in Fulton and DeKalb counties. 
To identify which precincts were affected by bomb threats, we cross-referenced news reports of threatened polling locations with official assignments of precincts to polling locations. 
%Each row in our dataset represents a precinct and contains the amount of votes per vote method and an indicator for whether or not the precinct experienced a threat. 
Historical data for the 2020 general election was downloaded from Clarity Elections, the commercial entity that stores Georgia's voting records.

Assembling the data for testing involved some nuisances. 
Fulton County often assigns multiple precincts to one polling location: a bomb threat at a single location may affect several precincts.
Thus bomb threats on different Fulton precincts can be statistically dependent.
Turnout data are reported by precinct, and thus there are multiple ways to process the data for testing. 
We addressed the dependence by aggregating turnout 
in the precincts at each polling location.
%One method to overcome this dependence is to perform statistical tests over polling locations rather than precincts by summing up the total votes per vote method over all precincts corresponding to each polling location. 
%One downside of this method is that it reduces the sample size and thus the power of the statistical test. 
%An alternative would be to treat the precincts corresponding to each location as one randomizable unit, thus keeping them in lock-step when building the null distribution for permutation testing. 
DeKalb only assigns one precinct to each polling location, so the preprocessing is simpler.

Some data gathered in November and December, 2024, was removed from the Georgia Secretary of State's website. 
When we built the initial data set, a few precincts in Fulton county were missing data on the breakdown of mode of voting. 
%Instead, the precinct would report the total number of votes for each candidate, but not the breakdown by vote method. 
Since that initial pull, data on mode of voting has disappeared for other precincts. 
Originally, data was missing for only one precinct in DeKalb, but data have since been removed for others.
Some data accessed through Clarity Elections has become inaccessible: previously valid web addresses to voting data now produce `403 Forbidden' errors. 
The website claimed that omitting mode of voting data for the 2024 election was necessary to protect voter identity in small precincts, but these data had
already been published.

\subsection{Using the 2020 Election as a Touchstone} \label{sec:why2020}
% increase of absentee/by mail ballots in 2020
% increase in 2024 as well compared to 2016
% take away - voter habits are dynamic and vary from precinct to precinct - must compare precincts to their previous behavior - does not make sense to compare to 2016
% expect positive change in in-person voting overall for 2024
To assess the effect of bomb threats and extended polling hours on turnout requires a touchstone: the turnout expected absent a threat (`control') is needed to compare the turnout of the precincts that received `credible' threats (`treatment').
It may be misleading to use precincts as controls for other precincts because political behavior may vary substantially by precinct. 
Hence, it is desirable to use historical data for a given precinct---turnout in a previous election---as a control.
Turnout in midterm elections is usually quite different from turnout in general elections, so it makes sense to compare 2024 to 2020 or 2016.
We are unaware of any polling-place bomb threats in Georgia in either of those years \cite{ica2021}.

Turnout by mode of voting in 2020 was quite different from that in 2016, because of the COVID-19
coronavirus pandemic. 
Emergency measures in 2020 expanded absentee voting, which increased substantially, especially among Democrats and in jurisdictions with higher rates of COVID-19 \cite{atkesonShouldVotebymailPerson2022, page2024democracy, herrnsonImpactCOVID19Election2023}.
Approximately 24\% of eligible Georgia voters requested an absentee ballot in the 2020 general election \cite{baccagliniGeorgia2020Election2020} compared to around 1\% in the 2016 primary \cite{brennanVote2020}. 
After the pandemic, many states continued to embrace absentee voting \cite{DemocracyMapsAbsentee}. 
Absentee voting decreased in Georgia in 2024, but a record-breaking number of ballots were cast early (including in-person `advanced' voting)~\cite{Raffensperger2024georgia}.
%Voting behavior has shifted away from the focus solely of in-person election day voting towards increases in early and absentee ballots since the COVID-19 pandemic. 
Hence, in-person 2024 voter turnout is more comparable to 2020 than to 2016, despite the pandemic.
However, the fact that 2020 was atypical needs to be taken into account in using it as a control.
For instance, all else equal, one might expect the in-person election day fraction of voter turnout in Georgia to be higher in 2024 than in 2020 since pandemic guidelines had been relaxed and access to absentee ballots was more restricted.

\subsection{Permutation Tests}
% - permutation testing overview
% + DeKalb randomized by location
% + Fulton randomized by location (pools over precincts) rather than precincts since threatened locations hosted multiple precincts
% - evaluation 
% - confidence intervals

A natural question is whether `credible' bomb threats suppressed in-person voter turnout, despite the extended voting hours. 
%Another question is whether the precincts that received threats were politically more polarized than other precincts, e.g., had a larger share of Democratic voters.
We address this question using 
\emph{permutation tests}, which rely on the fact that under the null hypothesis, given the orbit of the observed data under some group of transformations, any element of that orbit is equally likely to be observed.
\cite{pitman37a,pitman37b,pitman38} introduced permutation tests; the theory has evolved considerably since then (e.g., \cite{pesarin2010permutation}).
Monte Carlo methods for conducting permutation tests can be constructed in a way that yields conservative tests despite reliance on simulation
\cite{dwass1957modified,ramdas2023permutation,glazerStark25}. 

The null hypothesis is that
`received a credible threat' amounts to an arbitrary label that might as well have been assigned at random.
For example, in Fulton, five of 177~polling places received credible threats.
Under the null hypothesis that the threats had no effect, turnout would have been the same in any of the 177 whether it received a threat or not, so the label `received a credible threat'
might as well have been assigned at random to five of the 177~locations.
Implicit in the approach is the assumption
that turnout in a given precinct depends on whether it received a credible threat, but not on which other precincts received credible threats.
%Similarly, under the null hypothesis that the bombs threats did not specifically target polling places with an unusually high (or low) percentage of Democratic voters, the label `received a bomb threat' might as well have been assigned at random to five of the 177 locations.

\subsection{Test Statistic}
The null hypothesis `the polling places that received credible threats were selected as if at random' induces a probability distribution for any function of the precinct data.
What function of the data (\emph{test statistic}) should we use to test the hypothesis that extending polling hours offset any reduction in turnout that the bomb threats otherwise might have caused?

Since precincts vary in size, turnout, and voting habits---even in the absence of credible bomb threats---the raw number of in-person voters and the percentage turnout are not appropriate test statistics. 
The ratio of the number of ballots cast in-person 
in the precinct polling place on election day to the number cast early or absentee should be sensitive to changes in in-person turnout in a precinct, without being sensitive to overall turnout:
\begin{equation}
    m_j(y) := \frac{1+ \text{ in-person voters in precinct $j$ in year $y$}}{1+\text{early and absentee voters in precinct $j$ in year $y$}}.
    \label{eq:m}
\end{equation}
We call this \emph{the turnout statistic}
or \emph{relative in-person turnout}
(for year $y$).
However, that ratio may naturally systematically vary across precincts even without bomb threats, so it makes sense to normalize each precinct's ratio $m_j(2024)$ by the value of that ratio in a year in which there were no bomb threats.
We used 2020, for reasons given in Section~\ref{sec:why2020}.

We took the logarithm of the ratio of relative in-person turnout in 2024 to that of 2020 to change the scale to a percentage scale, resulting in the transformed data in Equation~\ref{eq:m_f}:
\begin{equation}
    \tilde{m}_j := \log \left ( \frac{m_j(2024)}{m_j(2020)}\right ).
    \label{eq:m_f}
\end{equation}
% metric interpretation
The permutation tests were based on the difference in the mean of $\tilde{m}_j$ for precincts that did or did not receive `credible' threats.
We explored tests using some other transformations of the data, such as raw voter turnout, but upon consideration, they 
seem further removed from the empirical question or more likely to confound the effect of threats with other factors.
%For this statistic, negative values would indicate fewer in-person votes compared to the previous election, while positive values signify increased in-person voting. 
%Values close to zero indicate similar behavior between the 2024 and 2020 general elections.

% mention that we did do the other tests but that they were not appropriate

\section{Results}
\label{sec:results}
This section presents exploratory data analysis
and the test results for the effect of `credible' threats on voter turnout (despite the extension of polling hours) in DeKalb and Fulton counties. 
The analysis primarily uses the statistic described in the previous section, for which positive values mean that in-person voting exceeded early and absentee voting.

\subsection{Exploratory Data Analysis}
\begin{figure}[h!]
    \centering
    \includegraphics[width=0.8\linewidth]{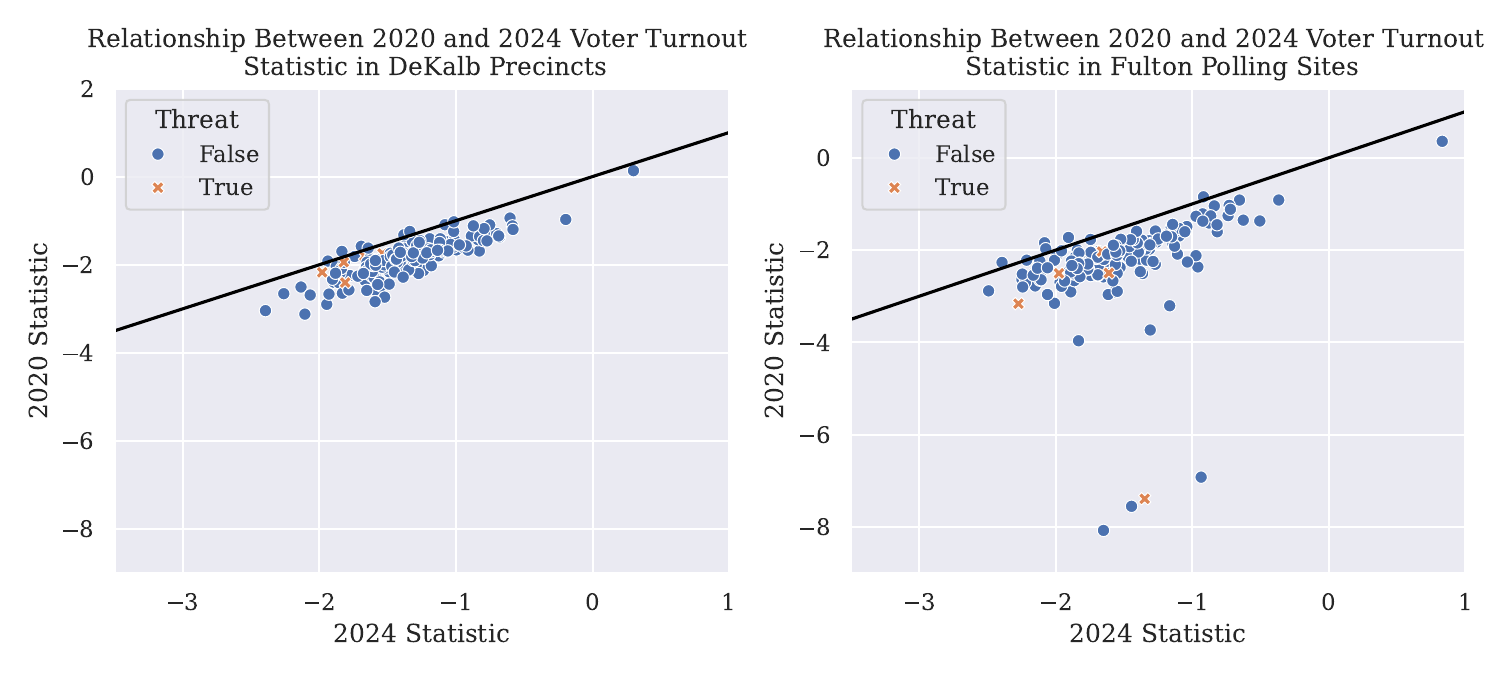}
    \caption{Scatterplot of 2024 and 2020 statistics for each precinct. The black line with slope~1 corresponds to precincts with the same fraction of in-person voting both years.}
    \label{fig:metric_scatter}
\end{figure}

% - Scatterplot of the metrics
% - geograohical plots 
Figure~\ref{fig:metric_scatter} plots the in-person election-day turnout statistic for 2020 and 2024. 
Precincts with a larger (relative to early and absentee) election-day in-person voter turnout in 2024 than in 2020 are below the line, which represents equal ratios in the two years. 
Almost every precinct had more early and absentee voters than in-person election-day voters, evident with a negative turnout statistic.
Most precincts in both DeKalb and Fulton had a higher relative election-day in-person voting in 2024 than in 2020.
Fulton has a few notable outliers at the bottom of Figure~\ref{fig:metric_scatter}: four locations had no in-person election-day voters in 2020.

\begin{figure}[h!]
    \centering
    \includegraphics[width=0.8\linewidth]{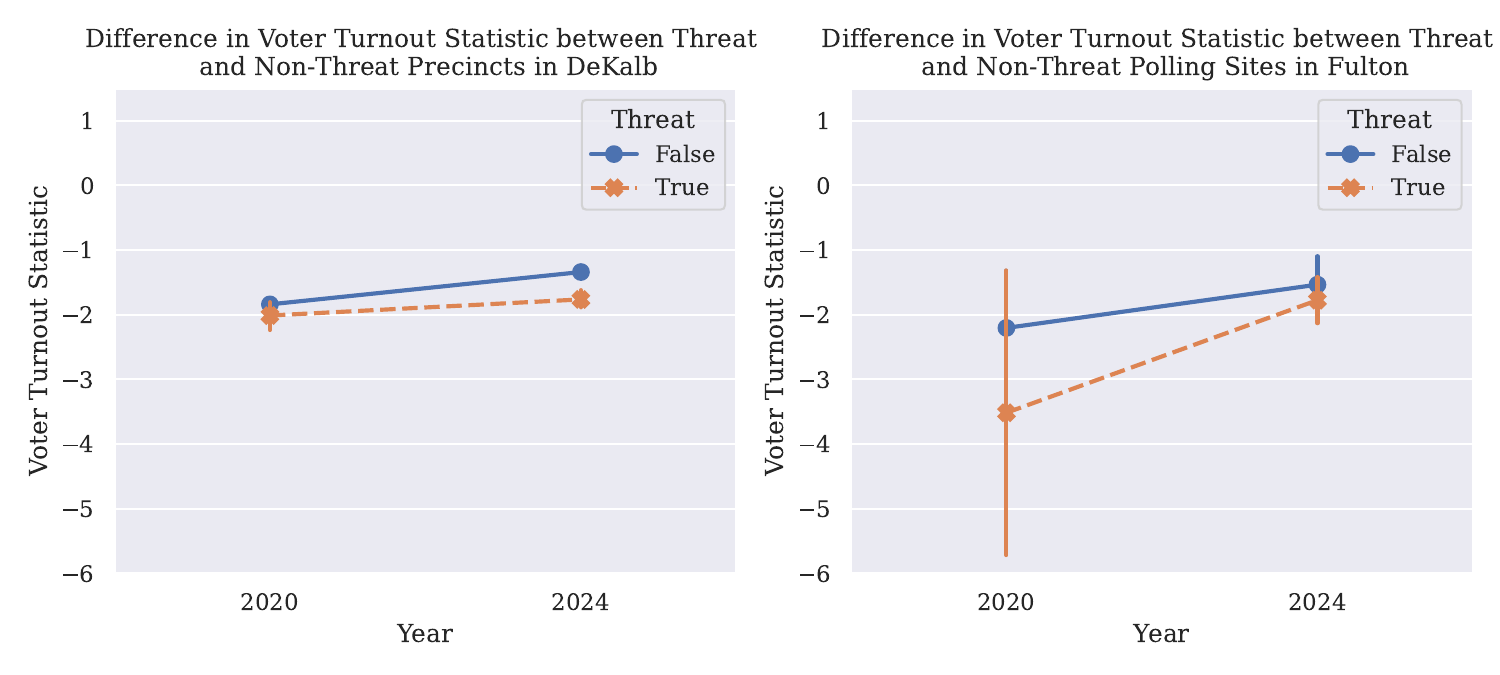}
    \caption{Change in the average relative in-person turnout
    for precincts with and without credible threats. 
    Six precincts DeKalb's 190 precincts received credible threats;
    five of Fulton's 177 precincts received credible threats.
    Vertical bars span $\pm 1$ standard deviation.}
    \label{fig:metric_averages}
\end{figure}

Figure~\ref{fig:metric_averages} summarizes the data in Figure~\ref{fig:metric_scatter} using averages and standard deviations. 
The mean of the turnout statistic for precincts with and without threats is shown for 2020 and 2024 for DeKalb and Fulton counties. 
The left panel of Figure~\ref{fig:metric_averages}, shows the gap between threatened and non-threatened precincts in DeKalb is larger in 2024 than in 2020, possibly due to bomb threats. 
However, in the right of Figure~\ref{fig:metric_averages}, Fulton county shows the opposite trend, with the gap between threatened and non-threatened polling locations shrinking in 2024. 
The standard deviation in 2020 is very large due to the outliers identified in Figure~\ref{fig:metric_scatter}. 

\begin{figure}[h!]
    \centering
    \includegraphics[width=0.7\linewidth]{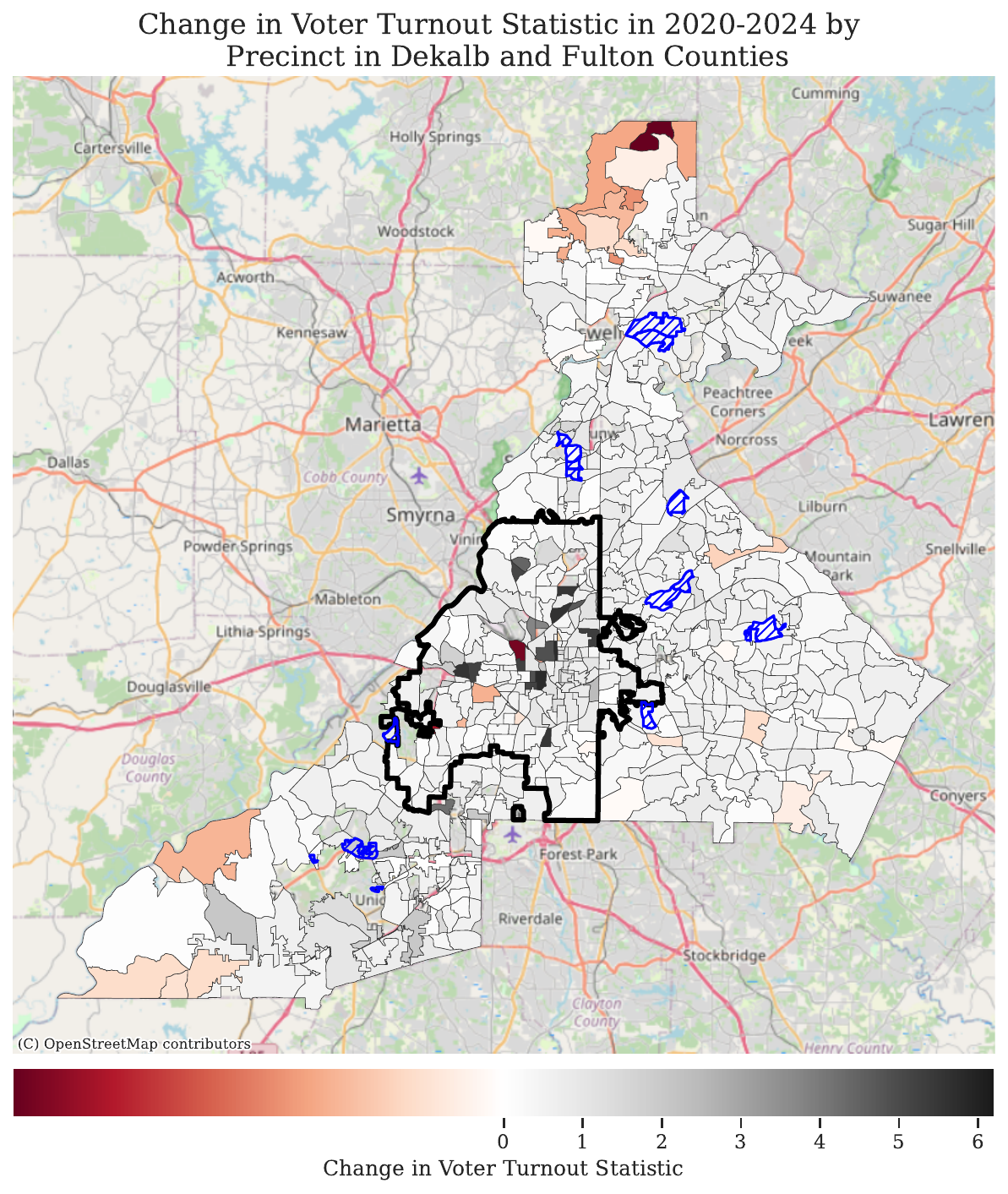}
    \caption{Change in relative in-person turnout between 2020 and 2024, by precinct. 
    Threatened precincts are crosshatched in blue.
    Positive values are increases.}
    \label{fig:geo}
\end{figure}

Figure~\ref{fig:geo} plots the distribution of the change in relative in-person turnout. 
The city limits of Atlanta are outlined in black, alongside both DeKalb and Fulton precincts, with threatened precincts cross-hatched in blue. 
DeKalb is in eastern Atlanta; Fulton covers the majority of Atlanta and regions to the south and north of the city. 
Most of the precincts in DeKalb with credible threats are shaded light white, indicating a small negative value. 
In Fulton, a few locations in the south with credible threats are darker: relative in-person turnout
increased.

\begin{figure}[h!]
    \centering
    \begin{minipage}{0.31\textwidth}
        \centering
        \includegraphics[width=\textwidth]{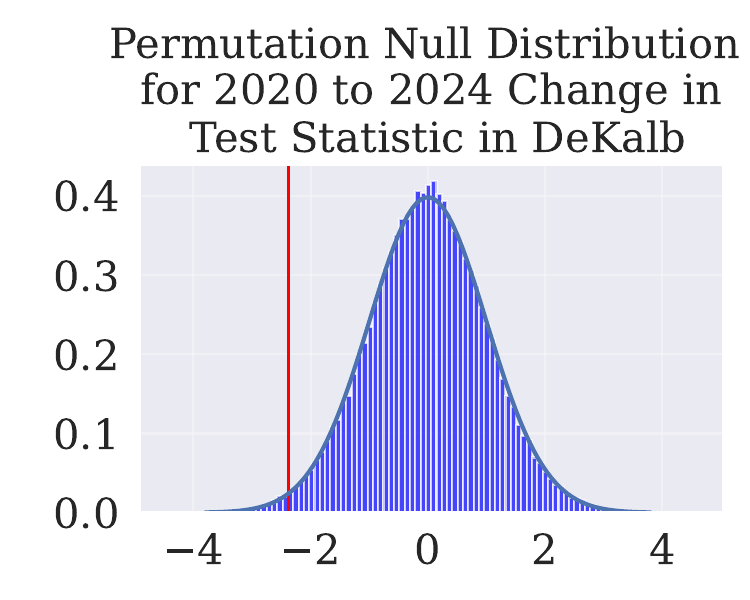}
        \caption{Distribution of log ratio of relative in-person turnout if credible threats have no effect after extending polling hours.
        Vertical bar marks the observed value for DeKalb County.}
        \label{fig:dekab_test}
    \end{minipage}\hspace{1.0em}
    %\hfill
    \begin{minipage}{0.31\textwidth}
        \centering
        \includegraphics[width=\textwidth]{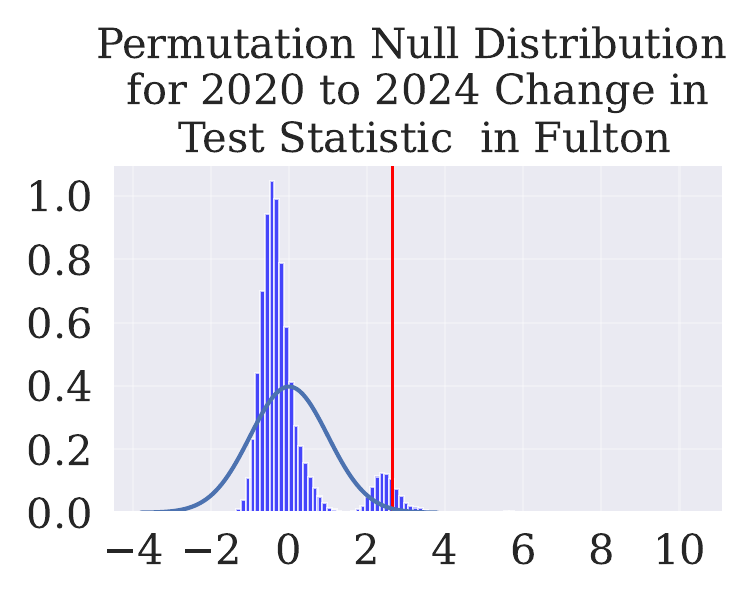}
        \caption{Distribution of log ratio of relative in-person turnout if credible threats have no effect after extending polling hours.
        Vertical bar marks the observed value for Fulton County.}
        \label{fig:fulton_test}
    \end{minipage}\hspace{1.0em}
    %\hfill
    \begin{minipage}{0.31\textwidth}
        \centering
        \includegraphics[width=\textwidth]{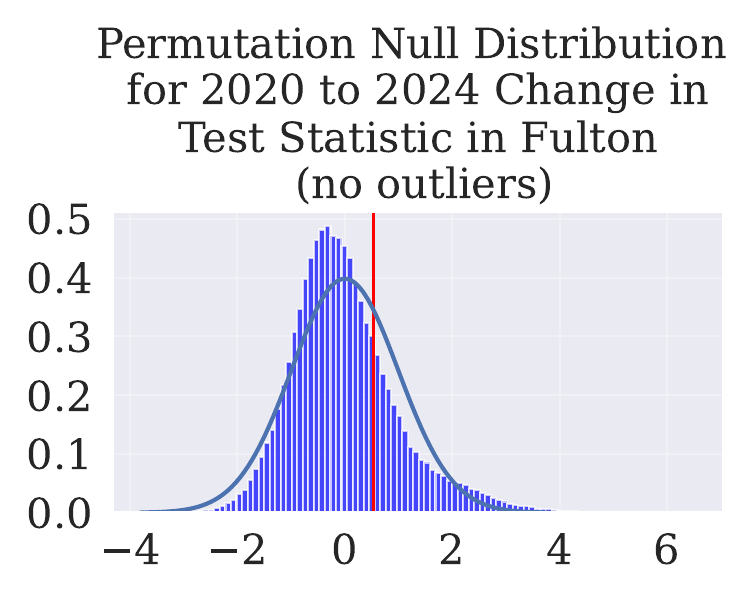}
        \caption{Distribution of log ratio of relative in-person turnout if credible threats have no effect after extending polling hours, omitting 4 precincts with no in-person votes in 2020.
        Vertical bar marks the observed value for Fulton County.}
        \label{fig:fulton_test_no_outlier}
    \end{minipage}
\end{figure}

\subsection{DeKalb}
% - threat effect
% - threat effect on outcome 
A permutation test of the hypothesis that the in-person election-day voter turnout in precincts with credible bomb threats was lower than that of other precincts in DeKalb county
yields a $P$-value of $0.01$, 
evidence that credible threats suppressed in-person voter turnout, despite the extension of polling hours.
The distribution of the log ratio of relative in-person turnout under the assumption that credible threats had no effect on turnout (after delaying the close of polls)  is shown in Figure~\ref{fig:dekab_test}, which plots a vertical line at the observed log ratio. 
The change in relative in-person voter turnout in the precincts with credible threats would be very unlikely under the null that the threats have no connection to voter turnout. 

\subsection{Fulton}
% - threat effect
% - threat effect on outcome 
A similar permutation test of the hypothesis that in-person election-day turnout is lower at polling locations (not precincts) with credible threats
in Fulton county did not find a statistically significant decrease ($P=0.965$).
In fact, the data provide some support for the contrary hypothesis that there was an increase in in-person voter turnout in those polling places (Figure~\ref{fig:fulton_test}). 
However, this result could be confounded by the fact that 32 Fulton precincts received bomb threats, while only 5 were closed:
if the `non-credible' threats suppressed turnout in the other 27 precincts, that could diminish or even reverse the sign of the apparent effect.
We were unable to identify the other 27 precincts to account for them specifically in the test.

Moreover, the four precincts in which no votes were cast in-person in 2020 increase the variance of the data and cannot possibly have a lower fraction of in-person turnout in 2024, potentially reducing the power of the test.
The $P$-value without those precincts is $0.755$ (Figure~\ref{fig:fulton_test_no_outlier}). 
%Figure~\ref{fig:fulton_test} displays the distribution of voter turnout statistics under the assumption that bomb threats had no effect, with the observed statistic in red falling in a highly likely region of the distribution. 

\section{Discussion } 
\label{sec:discuss}

While there is evidence that credible bomb threats suppressed turnout in DeKalb County, despite the extended polling hours, total turnout in  precincts closed by threats in DeKalb was over 75\%: 4,172 registered voters in the precincts closed for threats did not vote. 
Turnout in precincts in Fulton with credible threats was over 73\%; 5,794 registered voters in the closed precincts did not vote, for a total of 9,966 voters in the two counties.
Trump won Georgia by about 115,100 votes,
so even if the all those voters would have voted for Biden, Trump would still have won Georgia.
If turnout was affected \emph{only} in the precincts with `credible' threats, the threats did not change the outcome of the presidential election in Georgia.

\subsection{Implications and Limitations}
% implications:
% - connection to electino outcome
% - connect to historical impacts in these regions/voter intimidation/etc.
Election-day in-person turnout at polling places in DeKalb with credible bomb threats 
was lower by an amount that was 
noticeable and
statistically significant, despite the extension of polling hours.
In contrast,
extending polling hours in Fulton County may have compensated for the effect of the bomb threats
on in-person turnout.
This analysis  is complicated by a number of issues that limit power and may bias the conclusions:
\begin{itemize}
    \item The analysis assumes that threats that were not deemed `credible' had no effect on
    in-person turnout.
    Of 32 bomb threats in Fulton, only 5 were deemed `credible.' 
    If the other 27 reduced turnout, that would bias the analysis towards not finding an effect.
    \item The analysis assumes that whether turnout in a precinct was affected depends only on whether \emph{that} precinct received a 
    credible threat.
    If a threat at one precinct affects turnout at other sites, the $P$-values will be incorrect and the analysis will tend to understate the effect of the threats.
    It is not clear how to model adjacency effects without strong assumptions.
    \item There were no in-person voters in 2020 in 4 of Fulton's 177 precincts; in-person turnout at those sites cannot be lower in 2024.
    \item Fulton has some consolidated polling locations comprising multiple precincts, which reduces the effective sample size and therefore the power: the sample size is the number of polling locations rather than the number of precincts.
    \item Fulton was redistricted between 2020 and 2024: some precincts that existed in 2020 did not exist in 2024 and vice versa.
    Precincts that did not exist in both years could not be used in the analysis.
\end{itemize}
%Georgia's voting system requires election day voters to solely use electronic voting systems at their designated presinct. This rigid system resulted in the complete cessation of voting at polling places during the times they experienced threats.
%The threats had a statistically significant impact in DeKalb but not in Fulton.  

\subsection{Potential Mitigation} 
% - reactive vs resillient 
% - other methods than extending polling hours???
% - methods and pros/cons: allow to vote at other location, allow online voting (bad), late absentee ballots (executive order - not accept late ballots, 7pm after post closure), emergency paper ballots (), moving the devices outside, emphasis on early voting - spread out to decrease single point of failure (tension on potential for fraud vs one day event), back up polling locations (closed unless needed), mobile voting in vans (rapid response)
In the face of bomb threats, election administration in Georgia was \textit{reactive}---requiring judicial intervention that was not completely effective---rather than \textit{resilient}.
A system that allows voting to continue while a polling place is swept for bombs would be better.

Requiring all in-person voters to use BMDs creates a single point of failure for election-day voting. 
BMDs require electrical power, shade, and protection from the elements (and correct programming and configuration).
In contrast, voters could hand-mark ballots outdoors if necessary, using makeshift privacy screens.
Georgia law already provides for emergency hand-marked paper ballots, but they are rarely used.
Using hand-marked paper ballots as the default instead of universal-use BMDs would substantially improve election-day resilience---and would provide a much more trustworthy record of the vote.

Engineering system design might improve the robustness and resilience of elections.
There has been some work on pandemic-resilient elections, using simulation to estimate the impact of various failures on lines \cite{schmidt2022designing}.
Encouraging early voting (in person or by mail) decreases the number of voters who could be disenfranchised by election-day technology failures or bomb threats.
However, the more time between casting a ballot and tabulating the ballot, the greater the opportunity for ballots to be lost/removed, damaged, altered, injected, or substituted.
Thus there is a tension between security and resilience in concentrating voting on election day 
versus spreading voting out over time.

\section{Conclusion}
% restate findings
% connect back to event in georgia and election integrity
Extending polling hours at locations that were temporarily closed due to bomb threats 
did not entirely mitigate the suppression of in-person turnout at polling places closed due to `credible' threats. 
Precincts with `credible' threats in 2024 in DeKalb County, Georgia, had lower in-person election-day turn out than precincts without `credible' threats, relative to the turnout in 2020 in the same precincts, despite the fact that the polls were kept open late. 
However, the effect of `credible' bomb threats alone would not have changed the outcome of the presidential election in Georgia, assuming that such threats only affected turnout at the threatened polling places.

\section*{Acknowledgements}
This material is based upon work partially supported by the National Science Foundation Graduate Research Fellowship Program Grant No.~2146752 and
Grant SaTC~2228884. 
Any opinions, findings, and conclusions or recommendations expressed in this material are those of the authors and do not necessarily reflect the views of the National Science Foundation.

\bibliographystyle{unsrt}  
\bibliography{sources}

\end{document}